\documentclass[onecolumn,preprintnumbers,amsmath,amssymb,aps,floatfix]{revtex4-1}

\usepackage{amssymb,amsmath,graphicx,bm,textcomp,color,subcaption}

\begin{document}
\title{An all-optical
technique enables instantaneous single-shot
demodulation of images at high frequency}  

\author{Swapnesh Panigrahi$^{1}$}
\author{Julien Fade$^{1}$}
\email{julien.fade@univ-rennes1.fr}
\author{Romain Agaisse$^{1}$}
\author{Hema Ramachandran$^{2}$}
\author{Mehdi Alouini$^{1}$}

\affiliation{$^{1}$Univ Rennes, CNRS, Institut FOTON - UMR 6082, F-35000 Rennes, France}
\affiliation{$^{2}$Raman Research Institute, Sadashiv Nagar, Bangalore, 560080, India}
\date{\today}

\begin{abstract}
  High-frequency demodulation of wide area optical signals in a snapshot manner is a technological challenge that, if solved, could open tremendous perspectives in 3D imaging, free-space communications, or even ballistic photon imaging in highly scattering media. We present here a novel snapshot quadrature demodulation imaging technique, capable of estimating the amplitude and phase of light modulated from a single frame acquisition, without synchronization of emitter and receiver, and with the added capability of continuous frequency tuning. This all-optical setup relies on an electro-optic crystal that acts as a fast sinusoidal optical transmission gate and which, when inserted in an optimized optical architecture, allows for four quadrature image channels to be recorded simultaneously with any conventional camera. We report the design, experimental validation and examples of potential applications of such a wide-field quadrature demodulating system that allowed snapshot demodulation of images with good spatial resolution and continuous frequency selectivity, at modulation frequencies up to 500 kHz; no fundamental impediment in modulating/demodulating in the range 100-1000 MHz range is foreseen.

\end{abstract}

\maketitle

\section{Introduction} \label{sec:intro}

 Lock-in detection is an ubiquitous measurement technique, where the signal of interest is imparted a periodic variation at the source to distinguish it from the random noise that it acquires on its path to the detector. Extremely weak signals may be thus extracted by selectively amplifying, at the detector, the component at the modulation frequency. In the field of optics, light intensity modulation/demodulation techniques have been employed, for example, in telemetry, free-space communications, bio-medical imaging and viewing through scattering media. The enhanced immunity to noise and the massively increased transmission bandwidth due to  multiplexed modulation has been widely used in telemetry. On the other hand, a weakly modulated signal may be intentionally  cloaked in noise, with several  different messages being encoded at different frequencies. Despite the eventuality of an  eavesdropper demodulating a message at a certain frequency, the presence of multiple, possibly contradictory messages at different frequencies still provides partial secrecy (or at least discretion) of the communication. 
 
Another very important area where the modulation-demodulation technique plays a dominant role is in the imaging through complex disordered media. 
Optical inhomogeneities within the medium indeed cause random multiple scattering of photons, altering the normally ballistic transport of light into  diffusive transport, which strongly degrades the image-bearing capabilities of the light, hence resulting in turbid medium, and poor visibility, as in the case of biological tissues or in fog.
Imaging of objects hidden in such media can be achieved by the  extraction of the ballistic photons, that constitute a very small fraction of the total photons reaching the detector, but which retain the information of the source (direction, state of polarization, spatial and temporal modulation) and can lead to direct imaging through a turbid medium. On the other hand, indirect imaging of embedded objects can be performed by estimation of optical inhomogeneities in the turbid medium from the detected scattered photons. As a result, imaging through complex disordered media has been addressed using various techniques ranging from the efficient but costly time-gated  techniques \cite{ber93,sed11}, to the comparatively inexpensive polarization imaging \cite{ram98,emi96,fad14} and spatial modulation techniques \cite{Cuccia2009}. 
In this context, the temporal modulation-demodulation technique utilises the fact that the forward scattered ballistic light travels in a straight-line path within the medium, maintaining a phase relationship with the modulation of the source, while the scattered diffusive light has a statistical distribution of paths and hence loses the unique phase relation with the source, allowing its contribution to be filtered out for sufficiently high modulation frequencies. For instance, modulation frequencies in the
range of 10-100 MHz would meet such requirement for transport applications (or
for usual 3D range-imaging applications), whereas imaging in biological scattering tissues would require very high frequency operation in the GHz domain \cite{pan16}. As a result, modulation-based approaches in the radio-frequency (RF) range have been confined so far to point-wise detection configurations \cite{kim08,nit77}. 
Regarding two-dimensional ballistic-light imaging, existing techniques invariably require some form of processing at the receiver, either electronically, mechanically, or via software, increasing the complexity of the system, and often, the processing time. 
For example, electronic lock-in detection permits demodulation at one location at a time, necessitating a step-scan of the detector \cite{emi96}.
Software based approaches obviate the need for a step scan, but the requirement of  obtaining images in real-time restricts  the length of the time-series that may be recorded for demodulation, thus limiting the frequency of use \cite {ram98,sud16} well beyond the 100 MHz-GHz range.

Clearly, rapid techniques providing wide-field demodulation imaging are greatly desirable as they would not only permit real-time applications like navigation, but would also open up  possibilities for for 3D ranging and imaging, vibrometry, optical communications, and specialised scientific instrumentation. 
 Such imaging at high frequencies would be a
leap forward for imaging through turbid media, a field of interest that has bearing on vision through
opaque scattering walls \cite{wan91,ber12,kat14,kan15,sud16,bad16},
medical diagnosis \cite{ben93,tro97,Boas2001,ban16}, food quality
analysis \cite{lu16}, transport safety
\cite{Watkins2000,hau07,fad14}, underwater vision \cite{sch04} and
imaging through fog \cite{ram98}. 
 Progressing  from single-pixel
lock-in detection to  simultaneous demodulation over millions of pixels to achieve snapshot image demodulation  would bring to the realm  observation of  spatially distributed and fast physical
effects that remain otherwise undetectable.
However, demodulation of light at radio frequencies and higher is known to present several practical challenges like phase synchronization, timing jitters, snapshot operation and difficulty in frequency tuning that have only been partially addressed by the few existing laboratory
demonstrations of full-field demodulation, based on image
intensifiers, Time-of-Flight (ToF) sensors, lidar systems \cite{pay08, net08, han12, li14, muf11}. 
While fast intensity-modulated light sources are easily available, full-field demodulation of images at high frequencies  still awaits a viable solution. This calls for a  radically new approach to modulation-demodulation imaging in order to overcome technological impediments to  rapid, full-field imaging.  
   Here we propose and demonstrate a technique of imaging where  the demodulation at the receiver is performed {\it  optically} to obtain two-dimensional images {\it instantaneously from a recording of a single frame of an ordinary digital camera.}
This technique  is equally applicable to communications, cryptography and ballistic light imaging. As evidenced below, it offers several advantages in addition to the benefit of providing snapshot images.
   
\section{Results}\label{sec:ppe}
\subsection{FAST-QUAD : Full-field All-optical Singleshot Technique for QUadrature Demodulation}

We report here the first realization of an all-optical full-field instantaneous single-shot demodulation imaging technique - FAST-QUAD (Full-field All-optical Single-shot Technique for Quadrature Demodulation), compatible with high-frequency operation up to the RF range. This is achieved by performing the demodulation of the {\it intensity}-modulated light source(s) in the {\it polarization}-space  at the receiver. For that purpose, we exploit the Pockel's effect in an electro-optic crystal that introduces a phase difference between two orthogonal components of light that is proportional to an applied voltage. This effect occurs at very high speeds, with response times of a few picoseconds \cite{bor13}. As detailed below, suitable electrical excitation of the electro-optic crystal and orientation of birefringent/polarizing optical elements automatically resolves light into the quadrature components, and achieves the demodulation {\it obviating the need for phase synchronization with the source}. Final image integration on a standard camera is performed during an exposure time that is several orders of magnitude above the modulation period, and thus {\it a single frame captured by the FAST-QUAD camera provides the demodulated full-field image.}

 \begin{figure}[!ht]
        \centering
        \includegraphics[width=0.6\columnwidth]{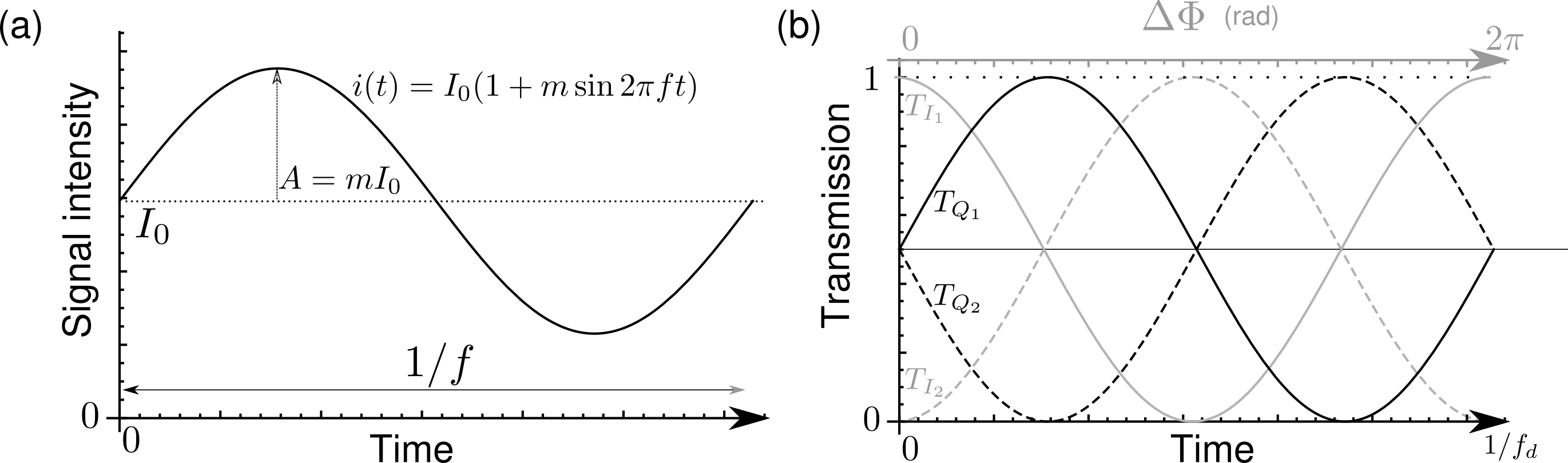}
        \caption{\textbf{Principle of quadrature demodulation
            imaging.} (a) At a given location (pixel) $(i,j)$ of the
          scene, the input light is assumed to be intensity modulated
          at frequency $f_{(i,j)}$ with modulation index $m_{(i,j)}$, over a mean
          (DC) intensity component $I_{0_{(i,j)}}$. (b) At each pixel of the
          detector, the incoming light is demodulated at frequency
          $f_{d}$ by product demodulation through four
          transmission-modulated optical gates along four quadratures
          ($T_{I_1}$, $T_{Q_1}$, $T_{I_2}$, $T_{Q_2}$). This is time-averaged on the camera over the exposure time of a frame, and the four intensity values
          ($I_{1_{(i,j)}}$, $Q_{1_{(i,j)}}$, $I_{2_{(i,j)}}$, $Q_{2_{(i,j)}}$) obtained at
          each pixel allow the average (DC) intensity
          $I_{0_{(i,j)}}$, amplitude $A_{(i,j)}=m_{(i,j)}\,I_{0_{(i,j)}}$ and phase
          $\varphi_{(i,j)}$ of light modulated at frequency $f_d$ to be
          simultaneously retrieved over the entire scene.}
        \label{fig_trans}
\end{figure}


Indeed, contrary to standard approaches that rely on temporal sampling of
modulated intensity signals (through image intensifiers, or specific
electronic chips such as ToF sensors),
 FAST-QUAD  requires no discrete temporal sampling of the received data. Instead, lock-in demodulation is performed continuously in time and simultaneously over the full spatial extent of the image. 
This is achieved by transferring the well-known quadrature demodulation (lock-in) principle to the optical domain and in a massively spatially-multiplexed way in order to handle image demodulation. 

A classic electronic lock-in detector multiplies an incoming signal that is modulated at a frequency $f$ by a sinusoid at frequency $f_d$ generated by a local oscillator.
The frequency and phase of this oscillator is tuned to obtain a coherent match with the weak incoming signal of interest. The product of the two is integrated over a length of time to average out all components except the one of interest. The phase matching step can be avoided when quadrature demodulation is performed, i.e, when the signal is demodulated by two local oscillators in quadrature (i.e., with a $90^\circ$ phase delay between each other) to obtain two demodulation channels ($I$ and $Q$ quadratures). Optically, the mathematical operation of multiplication of an incoming intensity-modulated light signal (or image) at a frequency $f$ with a local oscillator at $f_d$ can be achieved by passing the input light (or image) through an optical gate whose transmittivity is modulated sinusoidally at frequency $f_d$. Contrary to electronics, optical quadrature lock-in detection requires 4 transmission gates ($T_{I_j}, T_{Q_j}, j=1,2$) oscillating at $f_d$, with phases separated by $90\deg$ because the incoming light passes through optical transmission gates that non-zero transmittivity mean. This is shown schematically in Fig. \ref{fig_trans} for transmissions ($T_{I_1}=(1+\cos 2\pi f_d t)/2$, $T_{Q_1}=(1+\sin 2\pi f_d t)/2$, $T_{I_2}=(1-\cos 2\pi f_d t)/2$, $T_{Q_2}=(1-\sin 2\pi f_d t)/2$, as represented in Fig.~\ref{fig_trans}.b), these latter corresponding to four ``quadratures'' with respective phases $0$, $\pi/2$, $\pi$ and $3\pi/2$ radians. 

\begin{figure}[!ht]
        \centering
        \includegraphics[width=0.6\columnwidth]{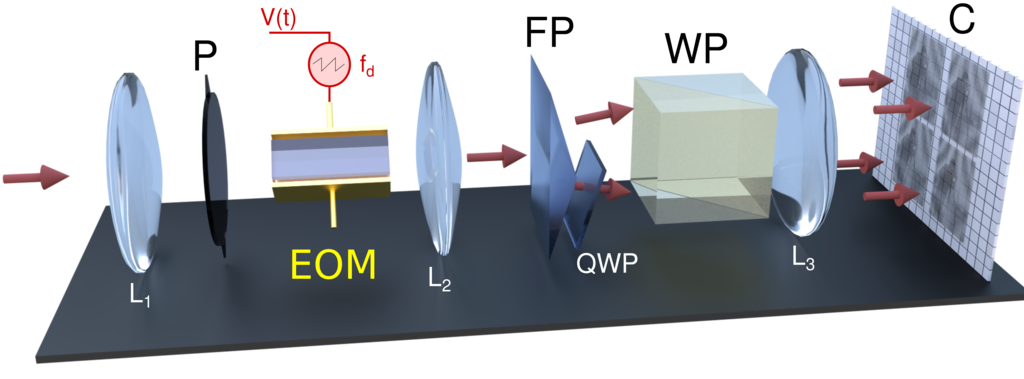}
        \caption{\textbf{Schematic of FAST-QUAD illustrating the principle of all-optical instantaneous full-field quadrature demodulation imaging.} The technique consists of the
            simultaneous recording, on a single camera frame of four
            sub-images, each of which is  the result of product
            demodulation, at frequency $f_{d}$, of the incoming light
            with specific phases ($0$, $\pi/2$, $\pi$ and $3\pi/2$
            radians). From these four sub-images, the average (DC)
            intensity and the modulation amplitude and phase maps can be
            obtained in a snapshot manner from a single acquired frame. To achieve this instantaneously and optically, the input light/image is passed  through a   lens $L_1$, a polarizer (P), and an
          electro-optic (EO) crystal driven by a
          high-voltage sawtooth signal at frequency
          $f_{d}$, which results in an optical phase difference
          excursion over $2\pi$ radians between two orthogonal
            polarization components. After collimation through lens
          $L_2$, the beam is split by a Fresnel bi-prism (FP), with the lower
          beam undergoing an additional $\pi/2$ radians optical phase
          shift passing through a quarter-wave plate (QWP). A
          polarizing Wollaston prism (WP) and a lens $L_3$ complete
          this 4-channel voltage-controlled sinusoidally varying optical
          transmission gate.
          \label{fig_3D}}
\end{figure}

Such parallel and instantaneous demodulation is achieved by use of a suitably designed arrangement of birefringent elements at the input of a standard low frame-rate camera (CCD or CMOS). The specific optical architecture employed for this purpose is illustrated in Fig.~\ref{fig_3D}. It comprises a polarizer (P), a quarter-wave plate (QWP) and
splitting/polarizing prisms (FP, WP), and a single electro-optic (EO)
crystal (e.g., Lithium Niobate (LiNbO$_3$)) with eigenaxes oriented at
$45^\circ$ from the input polarization axis imposed by the
  polarizer P. 
  A periodic sawtooth electric field at frequency $f_d$,  with sufficient excursion to ensure perfect sinusoidal optical transmission, (i.e., allowing a $2\pi$ radians excursion of the optical phase difference) is applied to the EO crystal.
As a result, this architecture does not require any dephasing or splitting electronic circuit. The $\lambda/4$ optical path difference arising in the QWP is
advantageously converted into a $\pi/2$ phase delay between optical
transmission curves $T_{I_{j}}$ and $T_{Q_{j}}$, {\it whatever be the
demodulation frequency $f_d$}, thereby offering huge tuning capabilities
of the demodulation imaging setup.
 The integration stage employed in conventional electronic quadrature
lock-in demodulation circuits is here simply and directly performed by acquiring a single frame on the standard camera (C) for a typical duration of several thousand modulation
periods. The optical arrangement is so designed that it allows the four quadrature images
($I_{1}$, $Q_{1}$, $I_{2}$, $Q_{2}$) to be simultaneously acquired on
the camera (see Fig.~\ref{fig_3D}). 

Finally, at each pixel $(i,j)$ of the scene, the average (DC)
intensity $I_{0_{(i,j)}}$, as well as the amplitude $A_{(i,j)}$ and phase
$\varphi_{(i,j)}$ of the light component modulated at frequency $f_d$ can be
retrieved from the four detected intensities, since
\begin{equation*}
I_{0_{(i,j)}}=\frac{I_{1_{(i,j)}}+I_{2_{(i,j)}}+Q_{1_{(i,j)}}+Q_{2_{(i,j)}}}{4},
\end{equation*}
\begin{equation*}
A_{k}=\sqrt{(I_{1_{(i,j)}}-I_{2_{(i,j)}})^2+(Q_{1_{(i,j)}}-Q_{2_{(i,j)}})^2},
\end{equation*}
\begin{equation*}
\varphi_{(i,j)}=\mathrm{atan}\Bigl[\frac{Q_{1_{(i,j)}}-Q_{2_{(i,j)}}}{I_{1_{(i,j)}}-I_{2_{(i,j)}}}\Bigr],
\end{equation*}
hence allowing instantaneous quadrature demodulation over the entire image from a single acquired frame.



\subsection{Experimental demonstration}\label{sec:results}

A prototype of a FAST-QUAD camera which implements the optical setup of Fig.~\ref{fig_3D} has been designed and built in order to validate and demonstrate the potentialities of the proposed full-field quadrature demodulation imaging approach. It includes a $40\times 2 \times2$~mm$^3$ lithium niobate (LiNbO$_3$) EO crystal and a low-frame rate
high-dynamic range sCMOS camera (Andor NEO sCMOS, $5.5$~Mpixels, $16$
bits) and was able to be operated up to few hundreds of kHz so far, limited by the bandwidth of the high-voltage amplifier available in our laboratory. The technical considerations and design of the prototype are reported in Materials and methods. The data processing pipeline and calibration procedure developped  to compensate for the  mechanical and optical
imperfections of this first prototype are given in Section~1 of Supplementary information.

The experimental validation reported below was performed with a green laser illumination ($\lambda=532$ nm), to limit the effect of the strong chromatic dispersion occuring in the optical components (especially EO crystal, prisms and QWP). A complete description of the imaging scenes considered in the remainder of this article is given in
Section~2 of Supplementary information.

\subsubsection{Proof-of-principle demonstration on an imaging scenario}
\begin{figure}[!ht]
        \centering
        \includegraphics[width=0.6\columnwidth]{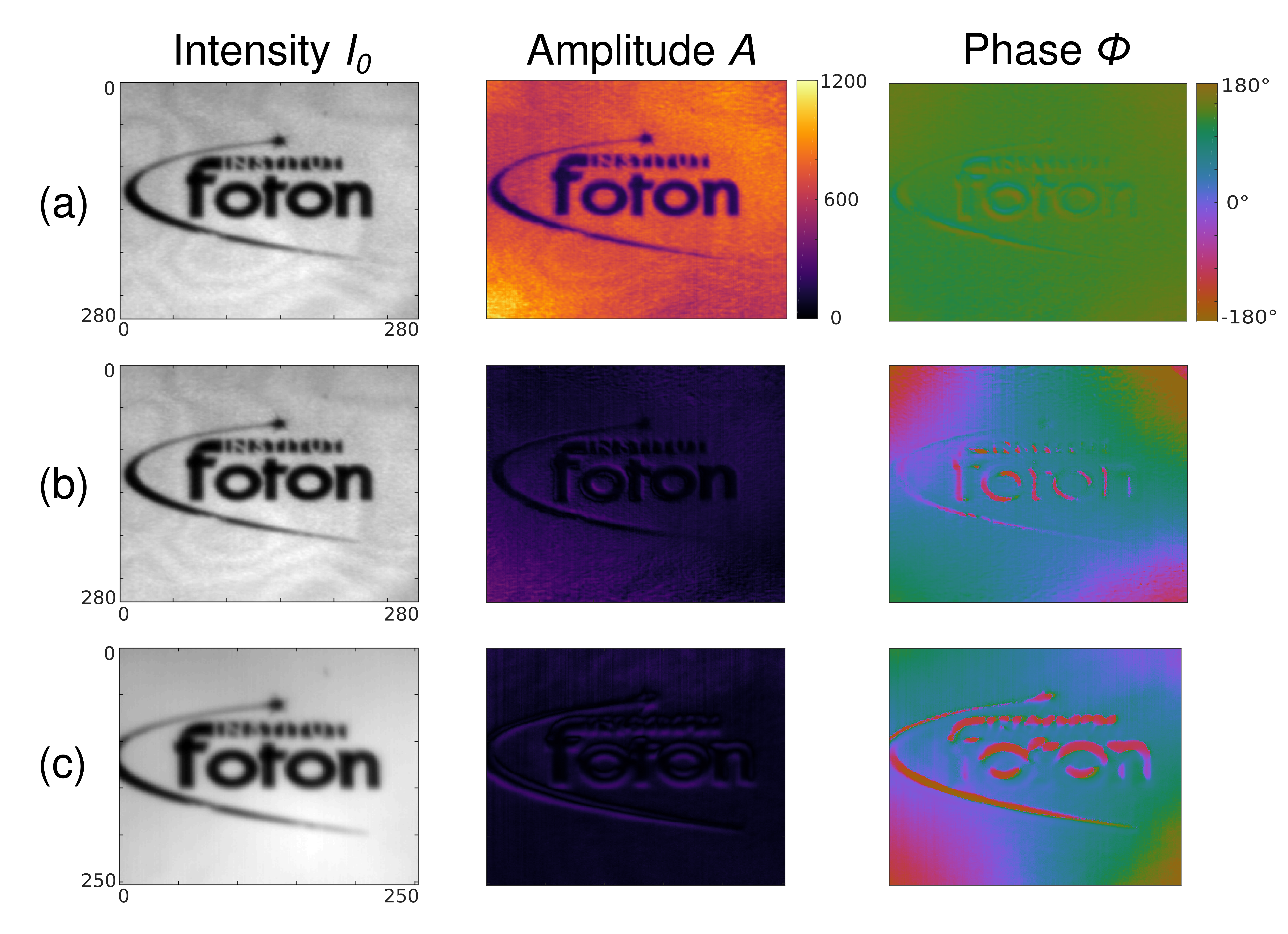}
        \caption{\textbf{Demonstration of snapshot image demodulation
            with FAST-QUAD.} 
            In this grid
of images, the left column contains the mean (DC) intensity of the source (the logo of Institut FOTON), the central column the amplitude map and the right column the phase map obtained by demodulation at frequency $f_d$ by FAST-QUAD. The first row corresponds to
the case where the scene is uniformly illuminated with laser light modulated at $f = f_d$, thereby resulting in strong demodulated amplitude signal and uniform phase map. The
second row corresponds to the source being modulated at $f = f_d + 10$~Hz, resulting in
almost null demodulated amplitude. The third row corresponds to the case where the source
is unmodulated. The demodulated amplitude is negligible in this case too, that mimics
illumination by spurious, ambient light.}
        \label{fig_logos}
\end{figure}

Instantaenous full-field demodulation imaging using FAST-QUAD has first been validated on a simple imaging scenario
where the source comprised a logo of the \emph{Institut Foton} (see
Fig.~\ref{fig_logos}) that was homogeneously illuminated by light modulated at $f=5$ kHz.

The demodulation frequency on FAST-QUAD  was first tuned to
the exact modulation frequency by setting $f_d=5$ kHz, and the corresponding results
are shown in the first row of Fig.~\ref{fig_logos}. The average (DC) intensity map of the incoming signal is  displayed on the left pannel, the amplitude map after demodulation in the central pannel and the phase map at the right. In this case, the modulation amplitude is well retrieved,
with the amplitude map showing a good homogeneity throughout the FOV
($280 \times 280$ pixels), and an appreciable spatial resolution. As expected,
the demodulated phase displayed in the right column leads to a fairly
flat estimated phase map. The quality of the demodulated images
demonstrates the efficiency of the calibration/processing algorithms
developed to compensate for the imperfections of the optical setup.

The demodulation frequency was then slightly detuned by setting
$\Delta f=f-f_d=10$~Hz, while maintaining the illumination intensity as illustrated in the left column of
Fig.~\ref{fig_logos}.b. In this case, the demodulated amplitude map in
Fig.~\ref{fig_logos}.b is almost dark, showing only a very
low-contrast residual image of the logo. The phase map is of course
not flat anymore, and would be expected to be irrelevant when the
receiver is not tuned with the emitter's frequency. The smooth
estimated phase pattern observed here is due to a residual lack of
correction of the phase mismatch across the isogyre pattern (see Materials and methods) that spreads across the raw images and which is corrected by calibration and
post-processing. It was also checked that the
demodulated amplitude was negligible when the scene was illuminated by
unmodulated white light, as shown in Fig.~\ref{fig_logos}.c, to
simulate strong ambient illumination. These first results obtained at
$f=5$~kHz hence demonstrate the ability of the FAST-QUAD to
efficiently demodulate an image in a snapshot manner with a good image quality and resolution.

\subsubsection{Evaluation of the frequency selectivity}

\begin{figure}[!ht]
        \centering
        \includegraphics[width=0.6\columnwidth]{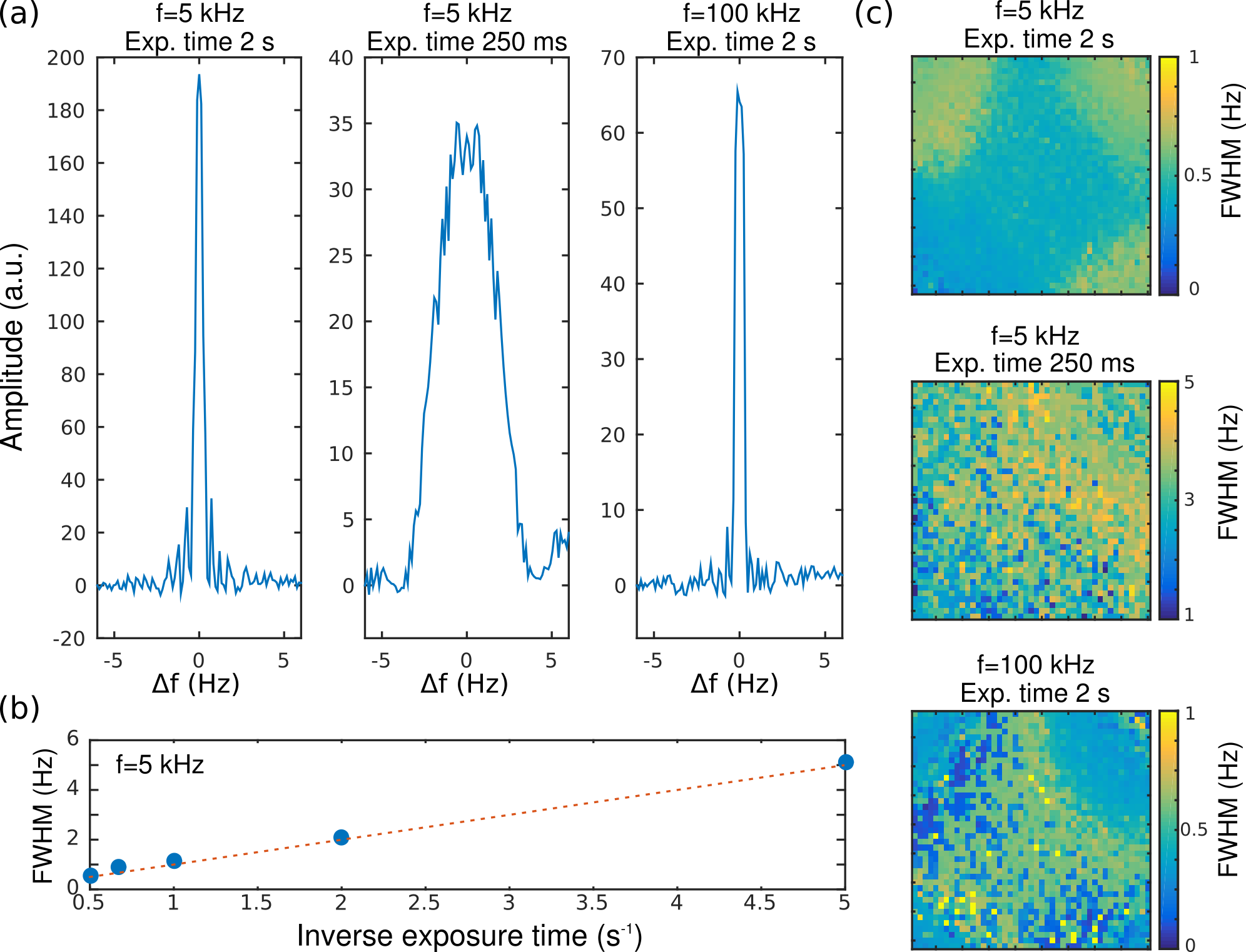}
        \caption{\textbf{Evaluation of frequency selectivity.}  (a)
          Demodulated amplitude averaged over $300\times 300$ pixels
          plotted as a function of frequency detuning
          $\Delta f= f-f_d$ for $f=5$~kHz and $2$~s exposure time
          (left); $f=5$~kHz and $250$~ms exposure time (center);
          $f=100$~kHz and $2$~s exposure time (right). (b) FWHM of
          the demodulation efficiency scales as the inverse exposure
          time. (c) Uniformity maps of the frequency selectivity (FWHM
          of the demodulated intensity as a function of $\Delta f$)
          over the entire FOV ($8\times 8$ pixels binning).}
        \label{fig_Qfactor}
\end{figure}

The demodulation frequency of FAST-QUAD can be continuously and easily tuned. To explore
the possibility of utilising this property to distinguish between images modulated at closely separated
frequencies, we investigated the frequency selectivity of FAST-QUAD on a homogeneous scene.
The average demodulated amplitude
across the FOV was evaluated as a function of the frequency detuning $\Delta f$.
A demodulation bandwidth of  $\sim 0.5$~Hz (FWHM) is obtained  at $f=5$~kHz and $f=100$~kHz  for a 2-s exposure time on the FAST-QUAD camera (Fig.~\ref{fig_Qfactor}.a); this increases to $\sim 2$~Hz (FWHM) when the exposure time is reduced to $0.5$~s.
Similar to usual lock-in
detection setups, the scaling of the selectivity with exposure time is
an expected result which is confirmed by Fig.~\ref{fig_Qfactor}.b,
where the FWHM of the demodulated amplitude is plotted as a function
of the exposure time for $f=5$~kHz. Next, the uniformity of the frequency
selectivity across the  FOV of the camera is analyzed in
Fig.~\ref{fig_Qfactor}.c, where the spatial evolution of the FWHM of
the demodulated amplitude is displayed using an $8\times 8$ pixels
binning. The selectivity of FAST-QUAD is found to be quite uniform across
the FOV, although the homogeneity is degraded with this first
  prototype when the global demodulation efficiency decreases with
higher modulation frequency or reduced exposure time.

\section{Discussion}

The previous section has shown the real-time image demodulation of . We now illustrate two important potential applications of FAST-QUAD: the reduction of clutter and image encryption.


  \paragraph{Frequency discrimination of several sources - the reduction of clutter:}
  
\begin{figure}[!ht]
        \centering
        \includegraphics[width=0.6\columnwidth]{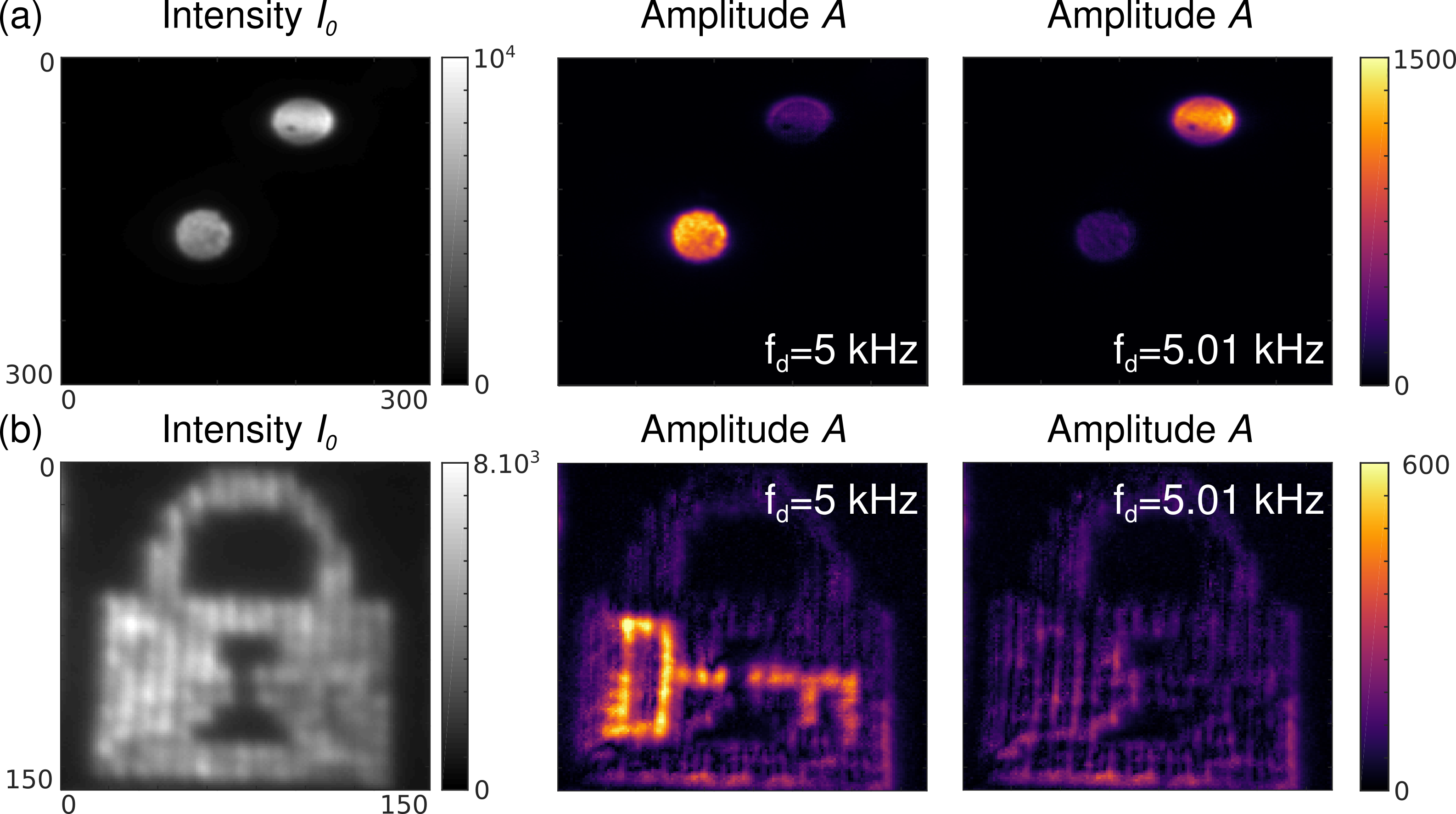}
        \caption{\textbf{Experimental illustration of potential
              applications of FAST-QUAD.} (a) Demonstration of frequency
            tuning: two laser spots of equal intensity and equal
            modulation index, but of distinct modulation frequencies
            (respectively $5.00$~kHz and $5.01$~kHz) are imaged and
            demodulated with FAST-QUAD (exposure time $2$~s). Left:
            Estimated average intensity map when demodulation
            frequency is set to $f_d=5$~kHz or $f_d=5.01$~kHz. Right:
            Modulation amplitude maps demonstrate that FAST-QUAD can
            be continuously ``tuned'' to any demodulation frequency,
            making it possible to actively select and discriminate
            emitters at different frequencies. (b) Demonstration of
            image encryption with modulated light: an
            intensity-modulated object (key) is concealed in an
            unmodulated background (lock), and the scene is imaged
            with FAST-QUAD (exposure time $1$~s). Left: The encrypted 
            image cannot be detected on a conventional intensity
            camera. Right: Modulation amplitude maps demonstrate that
            FAST-QUAD can retrieve the concealed image when $f_d$ is
            set to the exact frequency used by the sender.}
        \label{fig_2spots}
\end{figure}

  The imaging experiment presented in Fig. 6a, makes use of the continuous frequency discrimination capability of FAST-QUAD. Two objects (in our case, two disks) are illuminated by two independent intensity modulated sources with the same average intensity (see Fig.~\ref{fig_2spots}.a, left) but slightly different modulation frequencies ($5.00$~kHz and $5.01$~kHz). Tuning the demodulation frequency $f_d$ to one or the other frequency immediately results in a snapshot "filtered" image of the object modulated at that frequency in the amplitude map, demonstrating the frequency selectivity of FAST-QUAD (Fig.~\ref{fig_2spots}.a). Such discrimination capability opens up the possibility of using several sources at different modulation frequencies, permitting novel imaging applications like assigning
distinct frequencies to different classes of emitters, e.g., vehicles, road signs, landing areas, etc., and viewing each class in a de-cluttered fashion, by demodulating at their specific modulation frequency. As it requires no synchronization between the source and the receiver, the technique can be employed in the presence of relative motion between the source and receiver. This could, for example, help de-clutter the view of a pilot as he approaches for landing, and could aid in road, rail and other forms of navigation.

 \paragraph{Image encryption - Decoy image:}
 
 The second experiment, presented in  Fig.~\ref{fig_2spots}.b, utilises the
fact that FAST-QUAD provides frequency selectivity, and also requires no synchronization between the source and the receiver. It illustrates how a piece of secret information or an image (here, a picture of a key) could be embedded in decoy background (here, a picture of a lock) by the sender, and retrieved by the intended recipient equipped with FAST-QUAD, whereas it would go totally unnoticed by any other observer using a conventional camera. As the hidden object is intensity-modulated at high frequency (here, $5$~kHz) but with same average intensity as the
unmodulated background, a conventional camera does not allow for its detection (see intensity
map in Fig.~\ref{fig_2spots}.b (left)). On the other hand, a FAST-QUAD with the same exposure time makes it possible to detect the encrypted image efficiently (see amplitude maps in Fig.~\ref{fig_2spots}.b (center)),
when $f_d$ is set to the frequency used by the sender, without need for phase synchronization. A mismatch in the frequencies fails to reveal the embedded image (Fig.~\ref{fig_2spots}.b (right)).

In conclusion, we have proposed and demonstrated a novel all-optical technique for instantaneous full-field
demodulation of images from a single recorded frame of an ordinary digital camera. By means
of illustrative examples, we demonstrate the potential of this approach which has numerous
additional advantages like requiring no synchronization between source and observer, thereby permitting relative motion, continuous frequency tuning capability, compactness and portability. To the best of our knowledge, this proof-of-principle experiment is the first ever snapshot demodulation imaging in the DC to $500$~kHz frequency range, with the added capability of continuous frequency tuning.

These first encouraging results act as an incentive for achieving high-frequency snapshot quadrature demodulation imaging in the $10$'s MHz to $10$'s GHz range, which is no longer unrealistic. This would be a real breakthrough in the field of imaging in terms of applications, among which we can cite 3D imaging (potentially with range resolution below one millimeter), vibrometry, multiplexed free-space optical communications, automated vision, or spatially resolved lock-in detection at high-frequency. More importantly, this would also enable ballistic photons imaging through scattering media, with major applications in smart autonomous vehicles technologies and in biomedical diagnosis. Since the technique is based on the electro-optic effect (e.g., Pockel's effect with typical response times on the picosecond scale \cite{bor13}), the underlying concept of FAST-QUAD should remain valid even at those
very high frequencies. Technological bottleneck to operation at high frequencies would primarily be the fast application of high voltage signals across the EO crystal. This calls for an optimized optical design to limit the size of the EO crystal used while ensuring good image quality and resolution. Another challenge will be the generation of high-frequency voltage ramps to allow
for high-frequency operation. Lastly, conflicting requirements on the laser illumination linewidth imposed by good demodulation and speckle removal in the images will also be investigated in future developments.

\bigskip

\section*{Materials and methods}
\begin{figure}[!ht]
\centering
        \includegraphics[width=0.5\columnwidth]{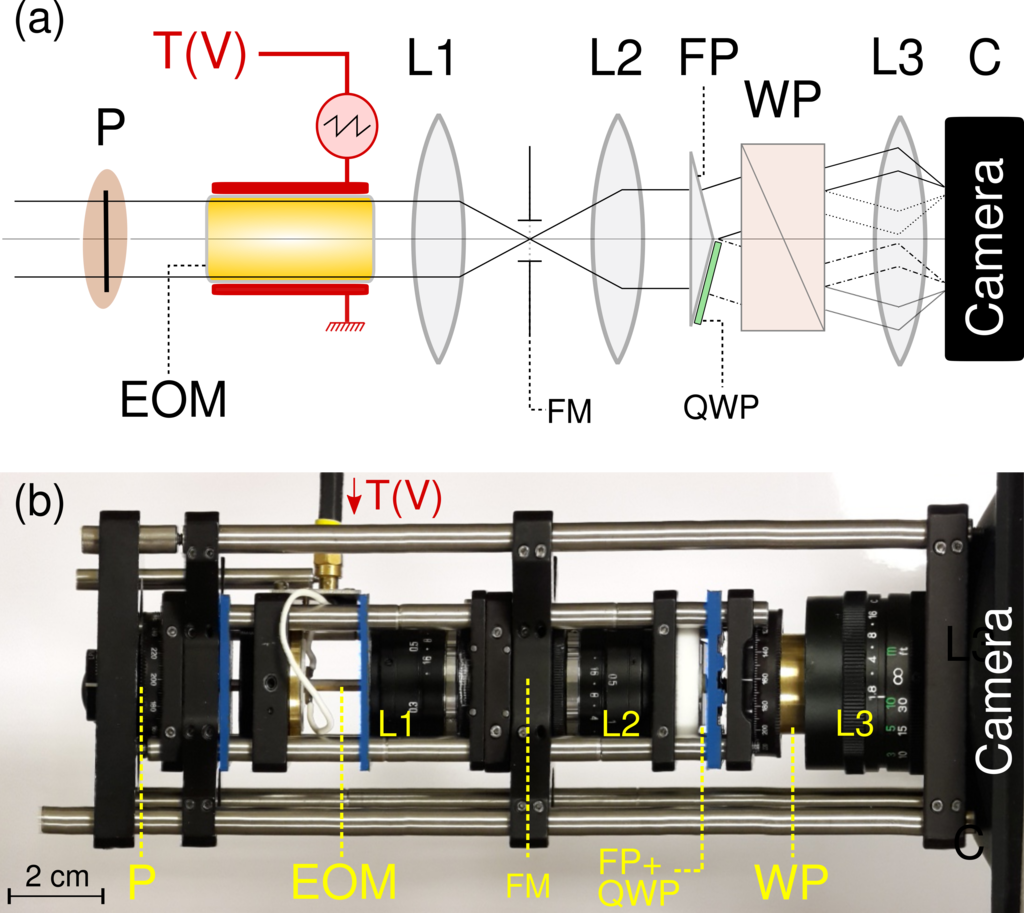}
        \caption{\textbf{(a) Schematic and (b) photograph of the prototype of FAST-QUAD.} Due to the dimensions of the EO crystal employed ($2\times 2 \times 40$~mm$^3$), it was optimal to have it positioned after the input polarizer $P$. This was followed by a focusing lens $L_1$, a field mask (FM) 
        at the intermediate image (that restricts the image spatial extent to prevent superimposition of the 4 sub-images on the camera), and thereafter a lens $L_2$ that recollimates the beam. A Fresnel biprism (FP) splits the beam into two, one part of which passes through a quarter-wave plate (QWP). Further propagation through a Wollaston prism results in 4 beams that are imaged onto the camera by means of lens $L_3$, providing the four quadrature images.}
        \label{fig_setup}
\end{figure}

The sketch and photograph of the optical setup used in this first
prototype of FAST-QUAD are given in Figs.~\ref{fig_setup}.a and \ref{fig_setup}.b respectively. This prototype is based on the general design described in Fig.~2 of the main article, and as stated there, the central component of the prototype is an electro-optic (EO) crystal which introduces
a controllable optical phase delay $\Delta \Phi(V)$ between two
transverse components of the light beam propagated through it, when a voltage $V$ is applied across it. Assuming
unpolarized incoming light with average intensity $I_0$,
and an ideal input polarizer $P$ (see Fig.~2 of the main article), the light
entering the EO crystal has intensity $I_0/2$ and is vertically
polarized, i.e., its Jones vector reads
$J_{in}=\sqrt{I_0/2}\, \bigl[ 1\ 0\bigr]^T$. The Jones matrix of the
EO crystal whose eigenaxes are oriented at $45^\circ$ from the input
polarization reads
\begin{equation*}
EO_{45^\circ}\bigl[\Delta \Phi(V)\bigr]=\begin{bmatrix}\cos \frac{\Delta \Phi(V)}{2}& i \sin \frac{\Delta \Phi(V)}{2}\\ i \sin \frac{\Delta \Phi(V)}{2}&\cos \frac{\Delta \Phi(V)}{2}\end{bmatrix},
\end{equation*}
whereas the action of an ideal Wollaston prism (oriented along the
input polarizer $P$) can be modeled with the following Jones matrices
of a vertical and an horizontal polarizer:
\begin{equation*}
\mathrm{WP}_H=\frac{1}{\sqrt{2}}\begin{bmatrix}1 & 0\\ 0&0\end{bmatrix}, \text{ and }\quad \mathrm{WP}_V=\frac{1}{\sqrt{2}}\begin{bmatrix}0 & 0\\ 0&1\end{bmatrix}.
\end{equation*}
An ideal Fresnel bi-prism simply entails a splitting of the
light intensity in half, with no effect on the beam
polarization. Lastly, the QWP with its
eigen-axes oriented at $-45^\circ$ from the input polarization
direction, has a Jones matrix equal to
$QWP_{-45^\circ}=EO_{45^\circ}\bigl[-\pi/4 \bigr]$. As a result, classical Jones calculus following the path of the beams across the optical setup
described in Fig.~\ref{fig_setup}.a leads to theoretical intensity
transmission functions for the four quadrature channels:
\begin{equation*}\begin{split}
  T_{I_1}&=\biggl|\frac{1}{2\sqrt{2}}  \mathrm{WP}_V\cdot EO_{45^\circ}\bigl[\Delta \Phi(V)\bigr] \cdot \bigl[ 0\ 1\bigr]^T\biggr|^2\\&=\frac{1+\cos \Delta \Phi(V)}{8},\\
  \end{split}\end{equation*}
  \begin{equation*}\begin{split}
  T_{I_2}&=\biggl|\frac{1}{2\sqrt{2}}  \mathrm{WP}_H\cdot EO_{45^\circ}\bigl[\Delta \Phi(V)\bigr] \cdot \bigl[ 0\ 1\bigr]^T\biggr|^2\\&=\frac{1-\cos \Delta \Phi(V)}{8},\\
  \end{split}\end{equation*}
  and for the "Q"-quadratures:
  \begin{equation*}\begin{split}
  T_{Q_1}&=\biggl|\frac{1}{2\sqrt{2}}  \mathrm{WP}_V\cdot QWP_{-45^\circ} \cdot EO_{45^\circ}\bigl[\Delta \Phi(V)\bigr] \cdot \bigl[ 0\ 1\bigr]^T\biggr|^2\\&=\frac{1+\sin \Delta \Phi(V)}{8},\\
  \end{split}\end{equation*}
  \begin{equation*}\begin{split}
  T_{Q_2}&=\biggl|\frac{1}{2\sqrt{2}}  \mathrm{WP}_H\cdot QWP_{-45^\circ} \cdot EO_{45^\circ}\bigl[\Delta \Phi(V)\bigr] \cdot \bigl[ 0\ 1\bigr]^T\biggr|^2\\&=\frac{1-\sin \Delta \Phi(V) }{8},
\end{split}\end{equation*}
which, on varying V, provide the transmission curves described
in Fig.~1 of the main article, thereby allowing spatially multiplexed lock-in
product demodulation of the four quadratures.

To attain an optical phase delay $\Delta \Phi(V)$  modulated at the demodulation frequency $f_d$ such that $\Delta \Phi(V)= 2 \pi f_d t\ \text{\textit{modulo}}\ 2\pi$, i.e., the
voltage $V$ applied across the EO crystal must be linearly
modulated along a sawtooth waveform with sufficient amplitude $V$.
The well-known theory of EO crystals provides the relationship
between $V$ and the optical phase difference. For instance, in the
case of lithium niobate as used in this prototype, one has
\cite{yar83}
$\Delta \Phi(V)= \pi (r_{13}n_O^3-r_{33}n_E^3) \ell / d \lambda \times
V$, where $r_{13}\simeq 10$~pm.V$^{-1}$ and
$r_{33}\simeq 30$~pm.V$^{-1}$ correspond to (Pockel's) electro-optic
coefficients of LiNbO$_3$ (approximate values in the visible range)
\cite{weis85}. The ordinary, respectively extraordinary, optical
refractive indices of LiNbO$_3$ are $n_O\simeq 2.32$ and
$n_E\simeq 2.23$ at $\lambda=532$~nm \cite{weis85}. To limit the
voltage amplitude to reasonable values (i.e., within the $100-200$ V
range), we selected a LiNbO$_3$ EO crystal (Moltec GmbH) of length
$\ell=40$~mm and with a $2\times 2$ mm$^2$ aperture section ($d=2$
mm). A high-voltage sawtooth waveform (with peak-to-peak amplitude of $124$ V
calibrated to provide exact $2\pi$ maximum optical phase difference
excursion $\Delta \Phi$) has been applied on the electrodes of the EO
crystal using a signal generator (Tektronix AFG3252C) and a
high-voltage amplifier (New Focus 3211 High Voltage Amplifier,
$\pm~200$~V, $0-0.6$ MHz bandwidth).

\begin{figure}[!ht]
        \centering
        \includegraphics[width=0.5\columnwidth]{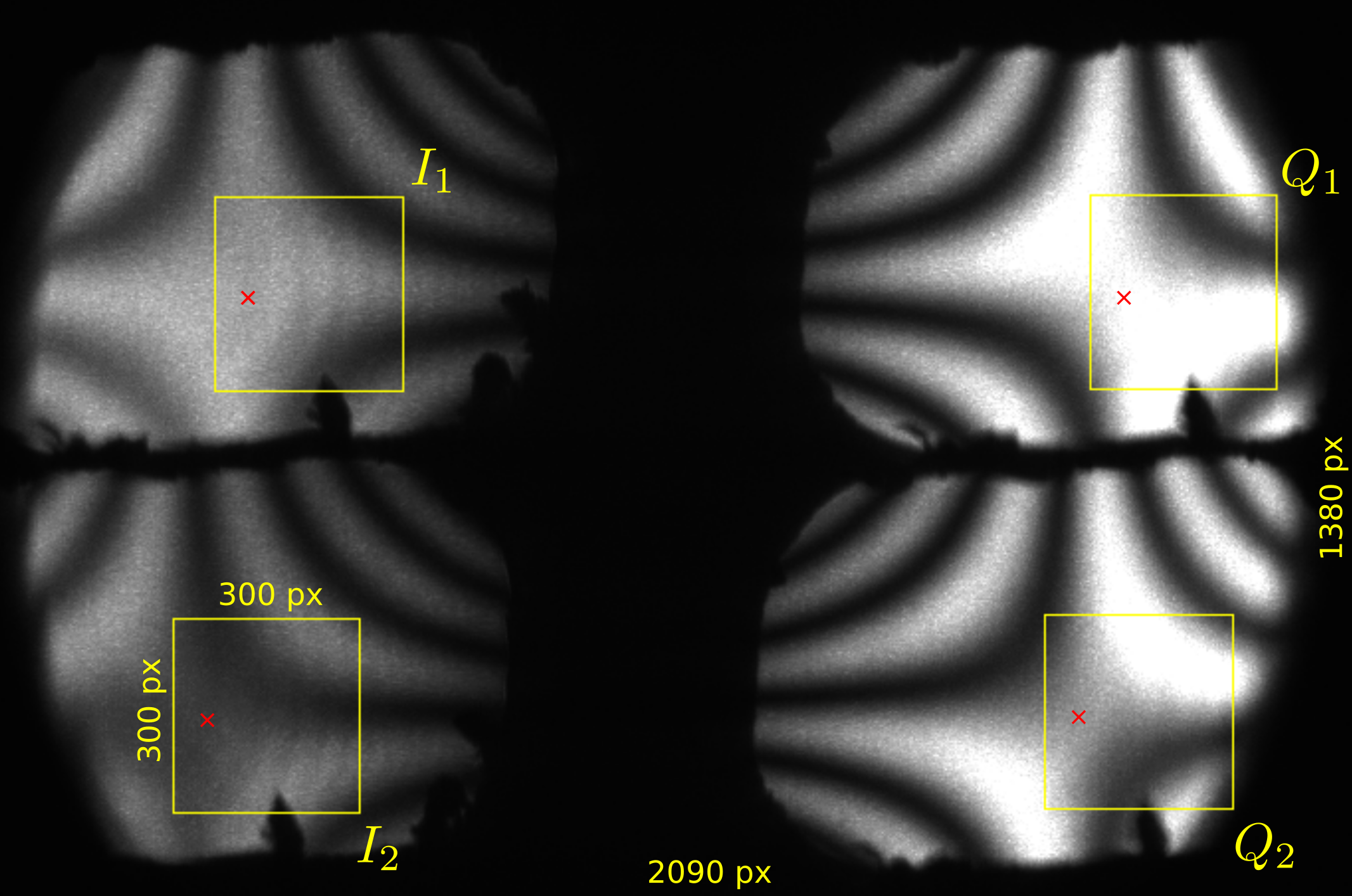}
        \caption{\textbf{Example of raw image acquired by the FAST-QUAD prototype
            with homogeneous illumination.} The four $300\times 300$
          pixels quadrature sub-images $I_1$, $I_2$, $Q_1$ and $Q_2$
          are delineated with yellow frames. The red cross indicates
          the position of the reference pixel used in section SI-4 to
          illustrate the quadrature mismatch correction algorithm
          implemented.}
        \label{fig_isogyres}
\end{figure}

Due to the importance of the ratio of the length and thickness/breadth
of the crystal, the best option in terms of FOV and resolution
consisted of placing the EO crystal just after the input polarizer $P$
(Thorlabs, LPVISB050), before the image was focused by lens $L_1$ in
an intermediate image plane where a field mask has been inserted to
restrict the spatial extent of the image so that the 4 sub-images do
not overlap at the camera (See Fig.~\ref{fig_setup}).  The light is
re-collimated through objective lens $L_2$ and split with a Fresnel
bi-prism (NewLight photonics, $160^\circ$ apex angle) and a Wollaston
prism (Melles Griot, $15.9$~mm, $5^\circ$ splitting angle). Lenses
$L_1$, $L_2$ are two $25$~mm focal length, $F/2.1$ camera objective
lenses, whereas $L_3$ is a $50$~mm, $F/2.8$ camera objective lens.  As
can be seen from Figs.~\ref{fig_setup}.a and \ref{fig_setup}.b, the
optical add-on offers relative compactness and ruggedness, which
qualities could be further improved in future developments with an
optimized optical and mechanical design.

A narrow bandwidth optical illumination was used with this setup
(green laser illumination at $\lambda=532$ nm) to limit the
detrimental effect on the image quality of the important chromatic
dispersion occuring in optical components (especially EO crystal,
prisms and QWP). As a consequence, the raw images acquired on the
camera are spatially modulated across the field-of-view (FOV) with an
interference pattern known as ``isogyre'', where the fringes
correspond to light paths in the crystal of equal optical phase delays
(birefringence) \cite{bor13}. Such a pattern is clearly visible on the
raw acquisition image example provided in Fig.~\ref{fig_isogyres}, but
pixel-wise, the four interference patterns obtained on the camera were
found to be advantageously in quadrature.  As a result, homogeneous
demodulated images were efficiently retrieved by implementing proper
calibration and processing of frames, as reported in Section~1 of
Supplementary information.

\section*{Acknowledgements}

The authors would like to thank L. Frein, S. Bouhier, C. Hamel and
A. Carr\'{e} for their technical help with the experiments. The
authors thank the Indo-French Center for Promotion of Advanced
Research (IFCPAR/CEFIPRA), New Delhi, for funding S. Panigrahi's PhD
and for supporting this research (RITFOLD project N$^\circ$ 4606).


\clearpage

\begin{center}
  \textbf{\large FAST-QUAD - A full-field, all-optical, single-shot technique for quadrature demodulation of images at high frequency: supplementary information}\\[.2cm]
  Swapnesh Panigrahi,$^{1}$ Julien Fade,$^{1,*}$ Romain Agaisse,$^{1}$ Hema Ramachandran,$^{2}$ and Mehdi Alouini$^1$\\[.1cm]
  {\itshape ${}^1$Univ Rennes, CNRS, Institut FOTON - UMR 6082, F-35000 Rennes, France\\
  ${}^2$Raman Research Institute, Sadashiv Nagar, Bangalore, 560080, India\\}
  ${}^*$Electronic address: julien.fade@univ-rennes1.fr\\
(Dated: \today)\\[1cm]
\end{center}

\setcounter{section}{0}
\setcounter{equation}{0}
\setcounter{figure}{0}
\setcounter{table}{0}
\setcounter{page}{1}

\section{Calibration procedure and processing of the frames}\label{sec:processing}
\subsection{General description of the calibration procedure and processing pipeline}
\begin{figure}[!ht]
        \centering
        \includegraphics[width=0.6\columnwidth]{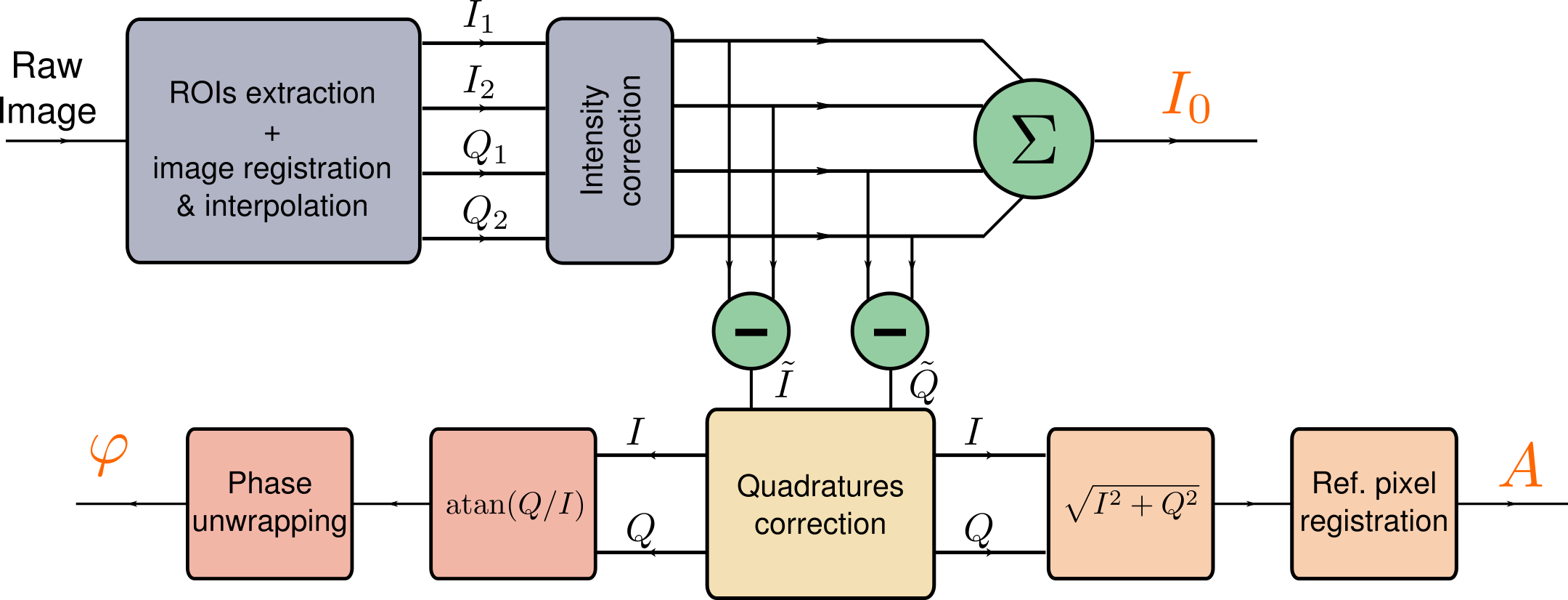}
        \caption{\textbf{Flowchart of the frames processing involved in FAST-QUAD.} The correction procedure  and the quadrature
          mismatch correction algorithm allow intensity $\hat{I}$, amplitude $\hat{A}$
          and phase $\hat{\varphi}$ maps to be retrieved in a snapshot
          way from a single frame acquisition.}
        \label{fig_flow}
\end{figure}

Optimal operation of the FAST-QUAD camera required several calibration and
processing steps to compensate for the mechanical and optical
imperfections of this first prototype. The flowchart of the data
processing involved to provide intensity, demodulated amplitude and
phase maps from the raw image acquired is sketched in Fig.~\ref{fig_flow}. Each
processing step is based on calibration data obtained from preliminary
experiments. Processing and calibration steps are briefly described
here. Since FAST-QUAD relies on the snapshot acquisition of the four
quadrature images on a single image sensor, a careful registration of
these four sub-images is mandatory to provide final images with
adequate resolution and to avoid estimation errors in the amplitude
and phase maps. A calibration pattern is imaged with FAST-QUAD,
allowing a first rough pixel-to-pixel registration to be performed,
leading to the four ROIs depicted in Fig.~7 of the main article. However due to the optical aberrations that differ from one
image channel to the other, it was necessary to implement a more
precise image registration, involving: (a) estimation of the
deformation maps during the calibration step on an appropriate
calibration pattern (regular dotted grid); and (b) image interpolation
of the $I_2$, $Q_1$, $Q_2$ images, so as to match the reference image
$I_1$. Full details on the calibration and interpolation procedures
implemented are not reported here for the sake of conciseness, as they
are strictly similar to the ones described in Ref.~\cite{fad14},
developed during prior work in the context of a two-channel
polarimetric camera.


Mainly due to imperfect mechanical alignment of the optical components
(polarizer, EO crystal, QWP, Wollaston prism), and to unbalanced
optical throughput in the beam-splitting components (Fresnel biprism
and Wollaston prism), the four sub-images retrieved from the raw data
acquired on the image sensor may have differences in average intensity
at each pixel, even in the case of perfect image registration.  Such
intensity mismatch is easily compensated for by applying appropriate
correction maps on each of the four quadrature images, after which
step the average intensity image can be directly estimated from the
four quadrature images by
$I_0=\bigl(I_{1}+I_{2}+Q_{1}+Q_{2}\bigr)/4$. The quadratures $I_1$ and
$I_2$ (respectively, $Q_1$ and $Q_2$) have been checked to be in
perfect opposite phase ($\pi$ radians phase shift) and to have
balanced optical throughput. As a result, two quadrature signals
$\tilde{I}=I_1-I_2$ and $\tilde{Q}=Q_1-Q_2$ are obtained by simple
numerical subtraction. However, due to residual optical misalignments
and experimental imperfections, the two signals
retrieved appeared not to be perfectly in
quadrature, which led us to implement a quadrature mismatch
calibration/correction algorithm inspired from RADAR signal
  processing \cite{chu81,noo99} (The  details of this quadrature mismatch
calibration/correction algorithm is reported in Section~2 of Supplement 1).

As sketched in Fig.~\ref{fig_flow}, the corrected quadrature signals
$I$ and $Q$ eventually allow the amplitude map $A$ to be obtained
after a final correction step, which is required to compensate for the
possible spatial inhomogeneity of the demodulation efficiency across
the FOV. Similarly, the estimated phase map $\varphi$ is retrieved
from the $I$ and $Q$ signals, after a phase ``unwrapping'' step to
compensate for the inhomogeneous phase distribution imprinted by the
isogyre pattern. The inhomogeneity map and the phase ``unwrapping''
pattern are also calibrated from initial measurements on a homogeneous
scene. The calibration procedure is not an easy task. Nevertheless, it
has in principle to be performed once for a given imager.

\subsection{\bf Quadrature mismatch calibration and the imaging
  algorithm}\label{sec:calib}

\begin{figure}[!ht]
        \centering
        \includegraphics[width=.6\columnwidth]{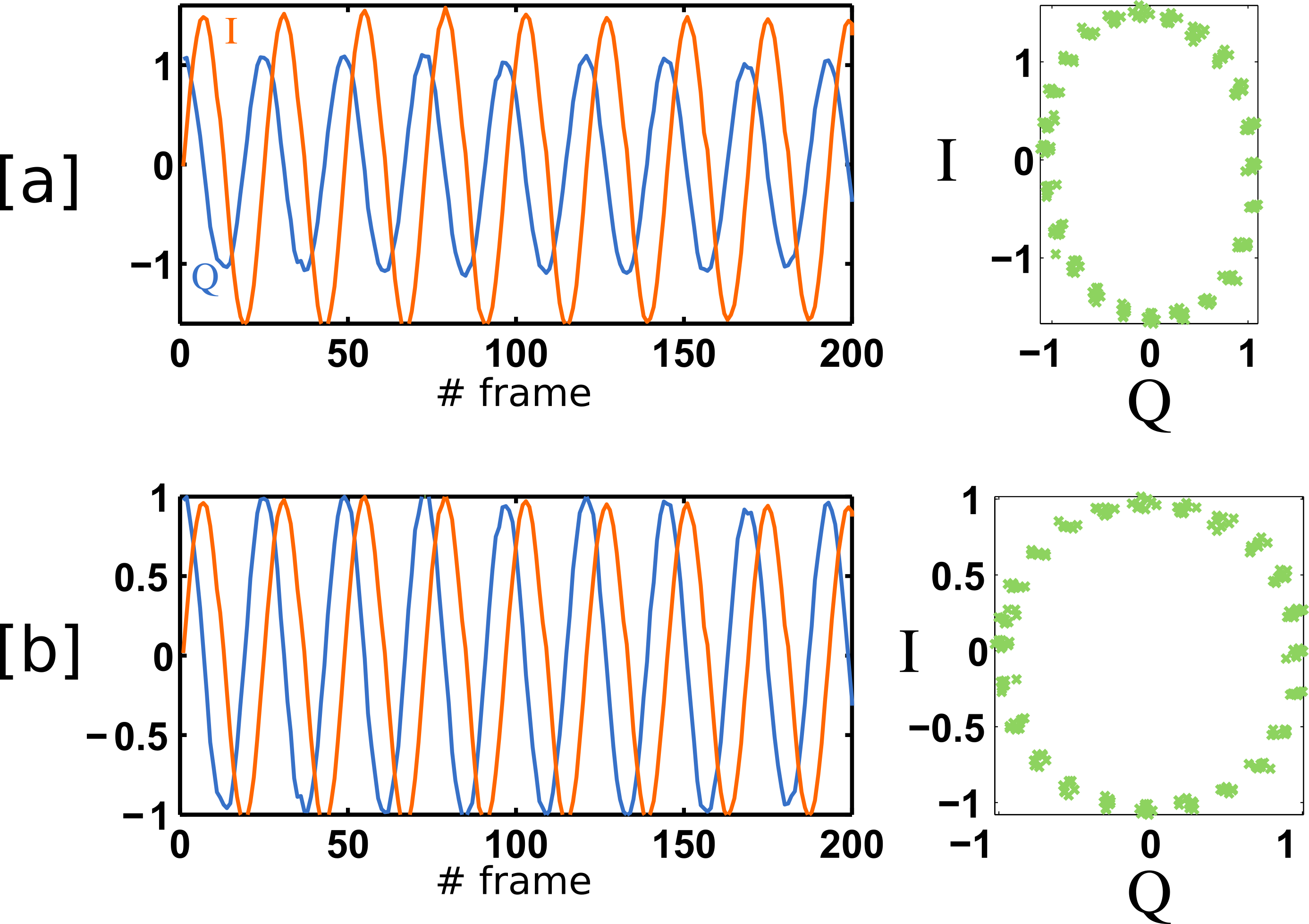}
        \caption{\textbf{Illustration of the quadrature mismatch correction
          algorithm.} (a)
          Original quadrature signals ($\tilde{I}$ in red, $\tilde{Q}$
          in blue) of the reference pixel (marked with symbol $\times$
          in Fig.~4 of the main article) as acquired by the
          camera over 200 frames ($350$~ms exposure time, $400$~ms
          sampling period) while EO crystal voltage is slowly varied ($0.1$
          Hz). The $I$-$Q$ representation (green dots) shows clear
          amplitude mismatch, and slight deviation from $90^\circ$
          phase (quadrature) between $I$ and $Q$ signals. (b) $I$ and
          $Q$ quadrature signals after correction showing equal
          amplitudes and perfect quadrature angle.}
        \label{fig_corrIQ}
\end{figure}

Due to experimental imperfections, the raw quadrature images obtained
from subtraction of two image channels, respectively
$\tilde{I}=I_1-I_2$ and $\tilde{Q}=Q_1-Q_2$, have small deviation from
being in quadrature phase. Moreover, a residual amplitude mismatch was
also observed between the two raw quadrature images $\tilde{I}$ and
$\tilde{Q}$. Such differences are clearly visible on the raw
quadrature data displayed in Fig.~\ref{fig_corrIQ}.a below, where the temporal evolution of the quadrature
signals $\tilde{I}$ and $\tilde{Q}$ are plotted as a function of time
for a reference pixel (denoted with symbol $\times$ in Fig.~4 of the main article), while applying a slowly varying ($0.1$~Hz)
voltage ramp on the EO crystal (with exposure time $350$~ms, sampling
period $400$~ms on an homogeneous scene).

The experimental data in this case can be modeled by writing the two quadrature
transmission functions at each pixel $k$ of the FOV as
\begin{equation}\begin{split}
    T_{\tilde{I}_k}&= T_{I_1,k}-T_{I_2,k}= \cos  (2 \pi f_d t+ \phi_k ), \text{ and}\\
    T_{\tilde{Q}_k}&= T_{Q_1,k}-T_{Q_2,k}= \alpha_k \sin (2 \pi f_d t+ \phi_k + \delta \phi_k),
\end{split}\end{equation}
where $\alpha_k$ accounts for the amplitude mismatch between the two
quadratures at pixel $k$, whereas $\delta \phi_k$ stands for phase
deviation from perfect quadrature between $T_{\tilde{I_k}}$ and
$T_{\tilde{Q_k}}$. It can be noted that a pixel-dependent common phase
factor $\phi_k$ has been included in the above equation, which models
the inhomogeneous phase distribution across the FOV, due to the
isogyre pattern. As mentioned in Section~4 of the main article, this phase distribution is
compensated in the last step of the frames processing during the phase
``unwrapping'' step to provide a correct estimated phase map. As a result, for
the sake of clarity in the following description of the quadrature
mismatch calibration and correction procedure, we will set $\phi_k=0$
without loss of generality.

The objective of the quadrature mismatch calibration described below is to estimate
the amplitude mismatch map $\alpha$, as well as the phase quadrature
mismatch $\delta \phi$ across the FOV, i.e., at each pixel
  $k$. Several calibration methods exist for correcting the intensity
and phase of a quadrature demodulator, especially for their
application in RADARs \cite{chu81,noo99}. We estimated the $\alpha$ and
$\delta \phi$ maps from calibration, using a time-series image
acquisition of $N=200$ raw frames, such as the one presented here in
Fig.~\ref{fig_corrIQ}.a for the reference pixel
  marked with a red cross symbol in Fig.~4 of the main article. The $N=200$
images temporally sample the amplitude response of the two quadrature
images $\tilde{I}$ and $\tilde{Q}$ over approximately $8$ periods,
with a spatially homogeneous constant (unmodulated) illumination on
the FAST-QUAD prototype, while the voltage applied on the EO crystal was slowly
varied ($f_d=0.1$~Hz).  Indeed, at a given pixel $k$ of the image, the
transmission can be written as a $N\times 2$ matrix
\begin{equation}
\mathrm{T}_k=\begin{bmatrix}(T_{\tilde{I}_k})_1 &(T_{\tilde{Q}_k})_1 \\ \vdots & \vdots\\ (T_{\tilde{I}_k})_N &(T_{\tilde{Q}_k})_N \end{bmatrix},
\end{equation}
whose covariance matrix yields
\begin{equation}
  \Bigl\langle \mathrm{T}_k^T \mathrm{T}_k \Bigr\rangle = \begin{bmatrix} 1 & \alpha_k \sin \delta \phi_k \\ \alpha_k \sin \delta \phi_k & (\alpha_k)^2 \end{bmatrix},
\end{equation}
showing that $\alpha_k$ and $\delta \phi_k$ can be easily estimated
from the calibration data. It can be noted that in the ideal case, the
covariance matrix should be equal to the identity matrix, denoting
perfect amplitude balancing and exact quadrature phase. 

Thus, one can estimate a scaling and a rotation matrix at each pixel
$k$ that can be applied to the observed data
$\mathrm{X}_k = [\tilde{I_k}\ \tilde{Q_k}]$ to balance the amplitudes
and obtain perfect quadrature signals. To calibrate these correction
matrices at each pixel, we follow a well-known method based on
singular values decomposition (SVD) and similar to \cite{noo99}. Briefly, the
SVD of matrix $\mathrm{T}_k$ is computed as
$\mathrm{T}_k = A_k \Sigma_k B_k^T $, and one can easily extract the
$2\times 2$ real unitary matrix $B_k^T$, as well as the two first
singular values stored in the $2\times 2$ diagonal matrix $\Sigma'_k$
(which is the $2\times 2$ sub-matrix of the $N \times 2$ matrix
$\Sigma_k$). As shown in \cite{noo99}, it suffices to apply a correction
matrix $R_k=B_k \Sigma_k'^{-1} B_k^T$ to the observed quadrature data
$\mathrm{X}_k$ to obtain corrected data $ \mathrm{X}_k R_k$ . The
efficiency of this correction algorithm can be checked in
Fig.~\ref{fig_corrIQ}.b where the corrected quadratures
$I$ and $Q$ are displayed, showing equalized amplitude and perfect
quadrature phase.

\section{\bf Description of the scenes being imaged}\label{sec:scene}

The scenes that were being imaged were illuminated with a $532$~nm
(Coherent Verdi) green laser. Acousto-optic modulators (AOM)
(AA~Opto-Electronic, MT80-A1-VIS) were used to obtain
intensity-modulated beams. Indeed, by supplying the AOMs with RF
signals whose amplitude was modulated at frequency $f$, the light
diffracted in the first diffraction order shows an intensity
modulation at frequency $f$ with modulation index $m\simeq100$~\%. A
sketch of the illumination setup is given in Fig.~\ref{fig_sphere} below. To obtain an image of an homogeneously
illuminated scene, the modulated beam was directed into an $8$~inches
diameter integrating sphere (Labsphere CSTM-US-800C-100R) which
created a uniform illumination field across its main circular aperture
of $5$~cm diameter. Note that during the experiments the sphere was
mechanically connected to an electrodynamic shaker (with frequency far
from the demodulation frequency, typ. $\sim$ 100 Hz) which created small vibrations of the
sphere, thereby limiting the detrimental effect of speckle on the
acquired images. To produce the images displayed in Fig.~6 of the main
article, a $4$~cm mask of the \emph{Institut Foton}'s logo was printed
on a transparency film and positioned in front of the aperture of the
integrating sphere. For the last experiments presented in Fig.~8 of
the main article, the two collimated laser spots (modulated at
distinct frequencies with two AOMs) directly illuminated a rotating
white diffuser (paper) used in place of the integrating sphere to
avoid speckle (Fig.~8.a). In Fig.~8.b, the collimated laser beams were used
  to illuminate white-dotted images of the lock (unmodulated) and key
  (modulation frequency 5~kHz), both printed on a transparency
  film. Two lenses were finally used to superimpose the final images
  of the lock and key on the rotating diffuser and create the scene of
  Fig.~8.b.

\begin{figure}[!ht]
        \centering
        \includegraphics[width=.5\columnwidth]{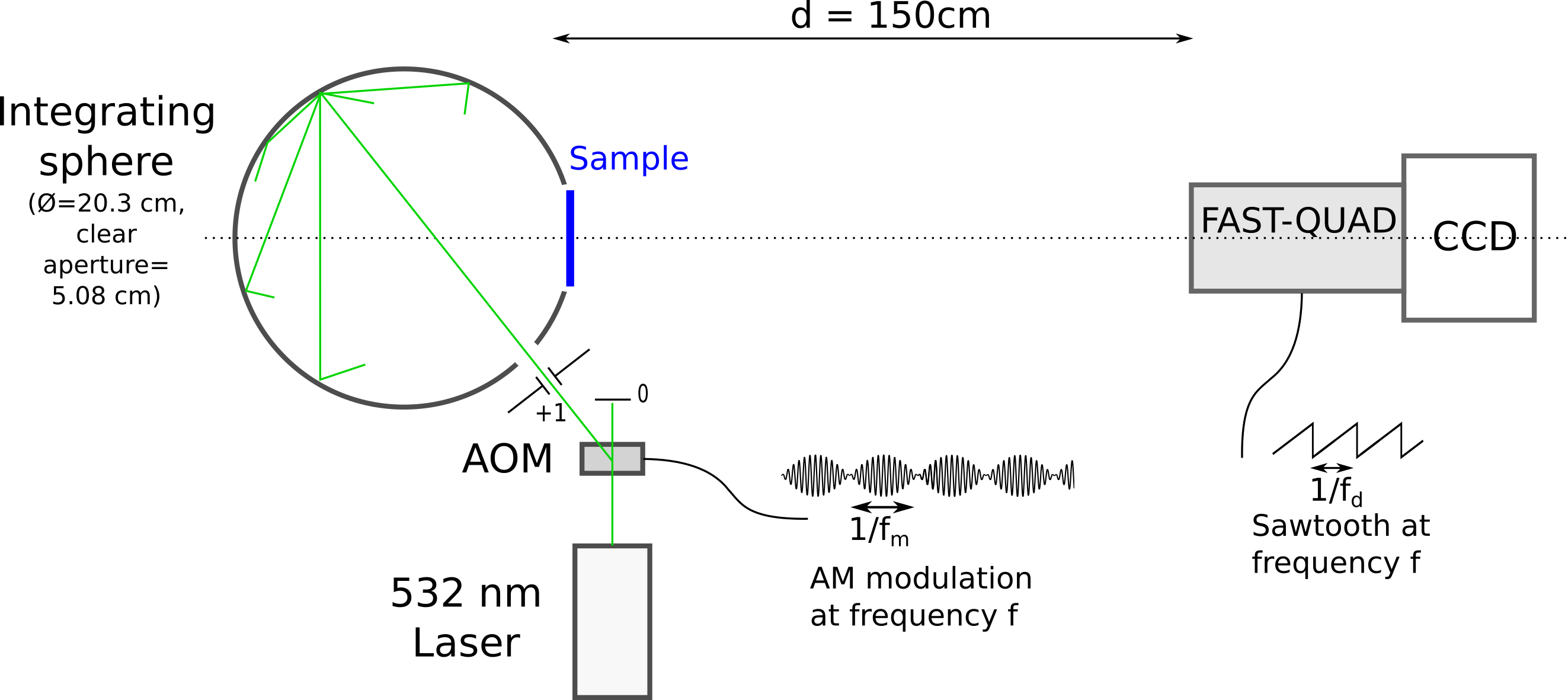}
        \caption{\textbf{Schematic of the experimental setup for imaging
          experiments with the FAST-QUAD prototype.}}
        \label{fig_sphere}
\end{figure}



\end{document}